\documentclass[11pt, oneside]{article}   	
\usepackage{geometry}                		
\usepackage{graphicx}				
\usepackage{amssymb}
\usepackage{amsmath}
\usepackage{epsfig}
\usepackage{epsf}

\title{Density vs distance for the DUNE beam from two recent geology density maps}
\author{Byron Roe\\
University of Michigan, Ann Arbor, MI 48109-1040}
\begin{document}
\maketitle
\vskip -75mm
\mbox{}\\
{U.M. DUNE technical note 2\\
DUNE-doc-1616-v2}
\vskip 60mm
\begin{abstract}
The densities passed through for neutrinos going from Fermilab
to Sanford lab are obtained using two recent density tables, crustal~\cite{crustal} 
and Shen-Ritzwoller\cite{Colorado}, 
as well as the values from an older table PEMC\cite{pemc}.
\end{abstract}
\section{Dividing up the path}
Twenty five points were selected taking equal intervals of latitude
and longitude.  However, the angle $\Delta\theta$ between the lines joining the endpoints
of these intervals to the center of the earth are not constant:
\begin{itemize}
\item{}
The size of the interval  $\Delta$longitude varies with latitude as cos(lat)
To get the correct angle $\Delta\theta$ it is necessary
to weight $\Delta$longitude by $\cos({\rm latitude})$.
\item{}
The earth is approximately an ellipsoid.  The radius in the North-South
direction is 6356 km and in the equatorial direction it is 6378 km.
$$(x/6378)^2 + (y/6356)^2 = 1.$$
Let the distance from the center of the earth to sea-level at a given
latitude-longitude value be $R$, the local radius.
Then $x = R\cos(lat);\ y = R\sin(lat)$
$$R = 1./\sqrt{(cos(lat)/6378.)^2+ (sin(lat)/6356.)^2)}$$
\end{itemize}

If we have a flat earth then then we would go from the initial height to
final height linearly with distance, $dist(i)$.
$${\rm flat\ height} = ({\rm endseaheight} *dist(i) +
{\rm startseaheight*(distfltosanford} - dist(i)))
/{\rm distfltosanford}.$$
The start height of the beam at Fermilab is 228.4 m above 
sea level and the end height of the midpoint of the detector at Sanford Lab is 159 m.
 The distance from Fermilab to Sanford lab $(FltoSl$) was taken as 1285 km. 

For the curved earth part starting and ending at sea-level with an arc of
$\theta$, imagine the angle of the arc goes from $-\theta/2$ to $+\theta/2$.
For 25 points, the midpoint, where the angle is zero, is to first order
given by the 13th point.  See Figure 1.  However, since the individual arcs between points
are not quite equal it is necessary to empirically slightly modify the midpoint from
13/25 to 13.02058/25. 
Some simplifications can be made.  Referring to Figure 1, it is seen that
$L = 2R\sin(\Delta \theta/2)$.  However, $\Delta \theta\approx 0.010$ radians, a very small value.
writing $L \approx R\Delta \theta$ introduces a negligible error in $\Delta L/L$ of $2.\times 10^{-5}$.
For this short segment the length of the arc and the length of the chord are essentially equal.

\begin{align} 
R^2 = (R-s)^2 + (L/2)^2\cr
(R -s)^2 = R^2 -(L/2)^2
\end{align}
Next look at $t$.  $t$ is not quite perpendicular to the straight line, but the error is small.
The fractional error in $t$ is zero at the center of the arc and incrases, approximately quadratically,
approaching a value of 0.5\% of the perpendicular distance by the end
of the arc, where $t$ is very small. 
\begin{align}
(R-s)^2 + (d-L/2)^2 = (R-t)^2
\end{align}
Substituting Equation 1 into Equation 2 
\begin{align}
R^2 -(L/2)^2 + (d-L/2)^2 = (R-t)^2\cr
(L/2)^2-(d-L/2)^2  = 2Rt -t^2 
\end{align}
We can ignore the $t^2$ term.
\begin{align}
t = [(L/2)^2-(d-L/2)^2]/(2R)
\end{align}
The distance above sea level is then given
by the sum of the flat height and the curved height.
There is an additional
effect called the geoid height, but it is very small, about 0.01 m for the Fermilab 
point and $-13.7$m for the Sanford lab point.

\begin{figure}
\includegraphics[height=16.5cm]{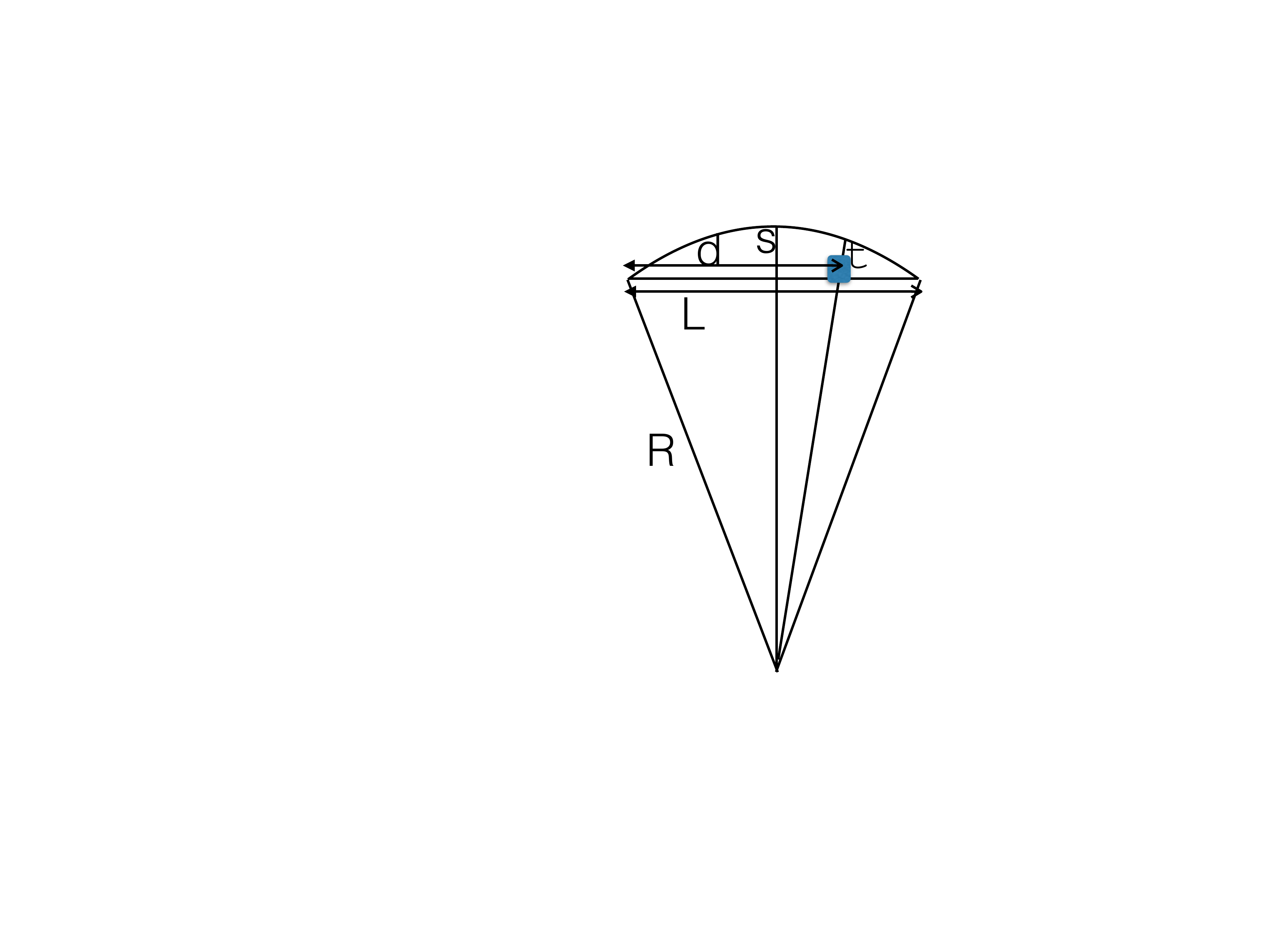}
\caption{ Figure to find the height of the earth surface above the straight line $L$ connecting
the sea level points at initial and final destinations.  $R$ = radius of circle, $s$ = distance
at the midpoint perpendicular to the straight line from the straight line to the circle (the sagitta) and
$t$ the distance from the straight line to the circle at a distance $d$ from the start. }
\label{Figure 1}
\end{figure}

Look at  Figure 2 to see how to get from point to point. Let $\Delta t$ be the contribution to $t$ from an
individual step and $\Delta\theta$ the change in angle.  Add $\Delta t$ to the previous $t$.
The angle with the
midpoint is $\theta_{\rm midpoint}$ - previous $\theta-\Delta\theta = \alpha$
and $\Delta t\approx \sin(\alpha)\times \Delta\theta\times R$.
The straight line distance from FL to SL is incremented by $\cos(\alpha)\times R\times
\Delta\theta$.
\begin{figure}[tbp] 
\includegraphics[height=25.5cm]{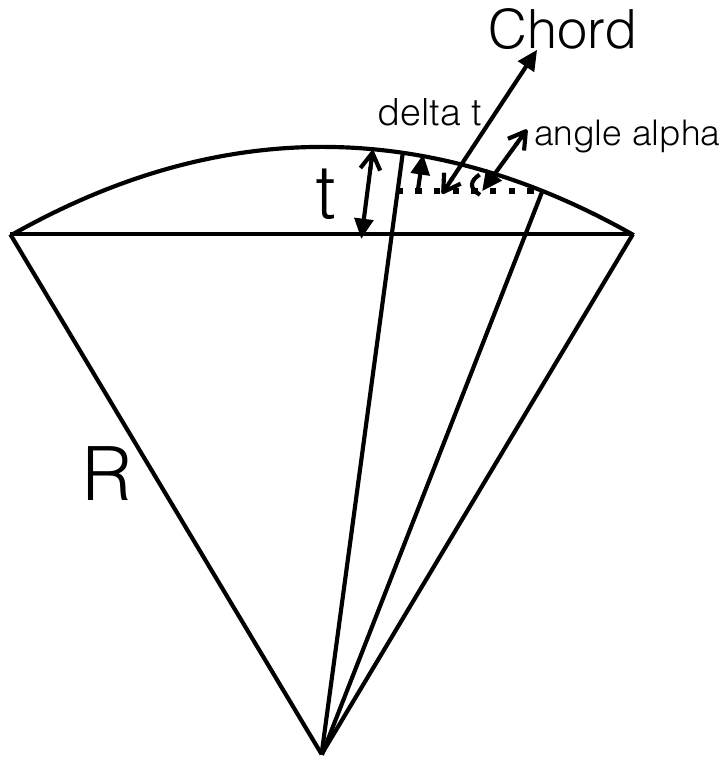}
\begin{center}
\vskip -85mm
\caption{ Figure to find the tilt ($\alpha$)  of the earth surface to the straight line connecting
the sea level points at initial and final destinations.  $R=$ radius of circle, $t =$ distance
from the straight line to the circle,  and $\Delta t=$ change in t from beginning
to end of segment.
}
\label{Figure 2}
\end{center}
\end{figure}
The density maps depend on the  depth of the beam below ground at the various points.
At Sanford Lab there are a number of hills and the beam ends up above sea level even though the
center of the detector is
close to 1470 m beneath the surface. The elevation at a given latitude and longitude can be 
obtained from a convenient web site\cite{FreeMap} and the difference between the elevation and the sea level
height of the beam is then the depth.  In general the elevation various smoothly except very
near Sanford Lab.  Had there been a lack of smoothness over a fair fraction of the path
it would have added considerable uncertainty to the density map.
\begin{figure}[tbp] 
\includegraphics[height=21.5cm]{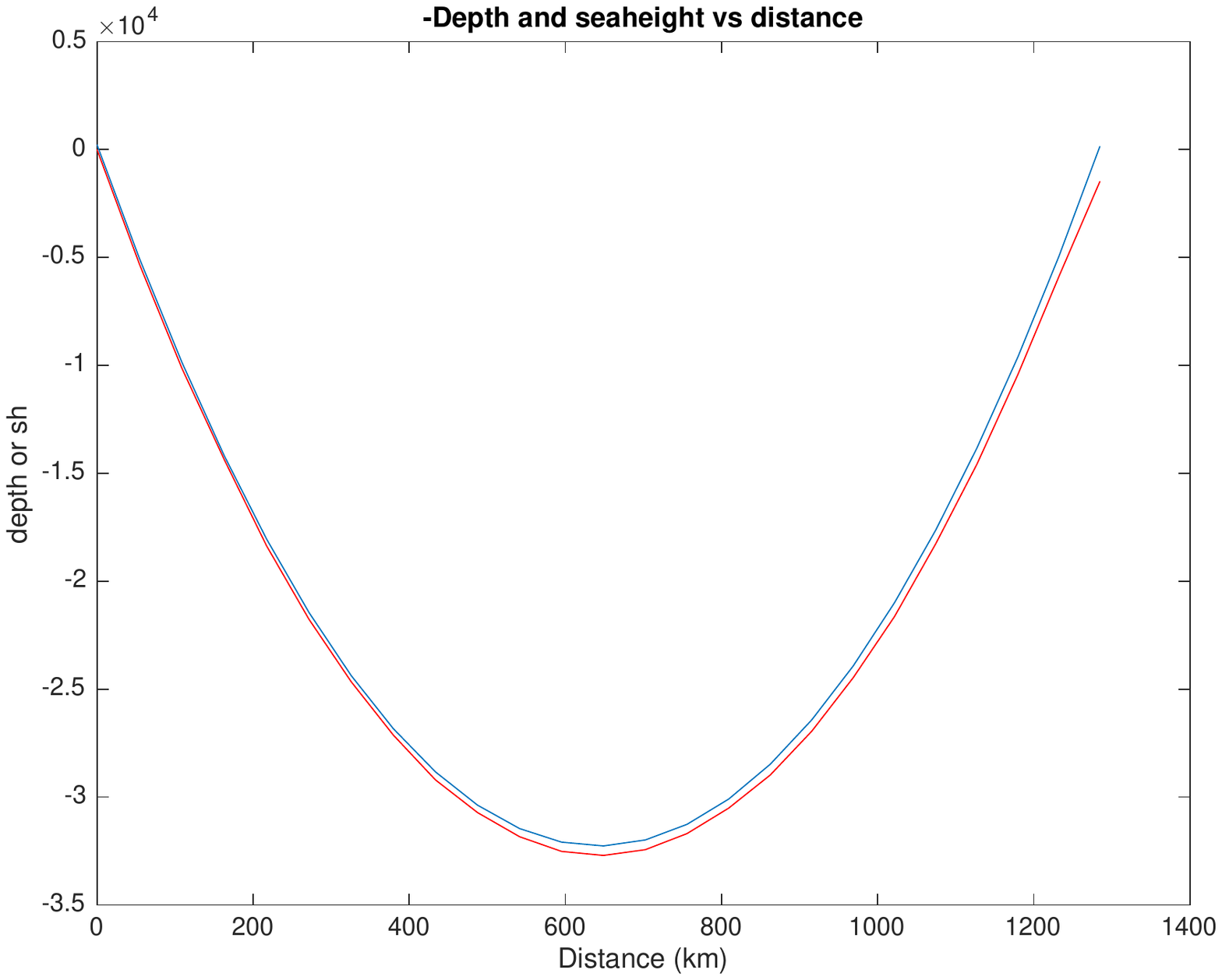}
\begin{center}
\caption{ Seaheight and -depth vs distance from Fermilab.  Blue is the seaheight; red is
the negative of depth.
}
\label{Figure 3}
\end{center}
\end{figure}
\clearpage

\section{The density maps}
Crustal~\cite{crustal} is a recent (2013) attempt to find the density of the earth as a function of latitude
and longitude. CRUST1.0 is an 8 layer model. The cells
average crustal structure over a 1x1 degree cells (about 110$\times$110 km), i.e. the grid
associated with the model is defined as the center of these cells.
If a model is therefore requested for a certain cell,  
the midpoint of this cell is to be used. For example, the model in the cell
5 to 6 deg latitude and 150 to 151 deg longitude should be inquired at
5.5 deg latitude and 150.5 deg longitude.  The map is based on ETOPO1 available from
NOAA's NGDC.

The 8 crustal layers:
====================
\begin{itemize}
\item{}1) water             
\item{}2) ice               
\item{}3) upper sediments   (VP, VS, rho not defined in all cells) 
\item{}4) middle sediments  \ \ \ \ "                  
\item{}5) lower sediments   \ \ \ \ \ \ "                    
\item{}6) upper crystalline crust                
\item{}7) middle crystalline crust             
\item{}8) lower crystalline crust  
\end{itemize}             

Although it is not needed here, a ninth layer gives V\_Pn, V\_Sn and $\rho$ below the Moho. 
The parameters below the Moho are determined using a modified version of the recent Pn model LLNL-G3Dv3 on continents and a thermal
model in the oceans.  

V\_Pn and V\_Sn are the compression (primary) and the shear 
(secondary) wave velocities of sound
in the medium.
The
 model is defined from 89.5 to -89.5 deg latitude and -179.5 to 179.5 deg
longitude. 

Comments:  Density is in gm$/$cm$^3$.  Our longitude (W) corresponds to negative values here.
For a given latitude and longitude, the Crustal supplied program getCN1point 
gives V\_PN, V\_SN, $\rho$
and bottom of each layer.
For all maps in this note, the depth, not the sea-level height is used in the maps.

The program getCN1point asks for input and then produces a set of densities and layer
bottoms:
\begin{obeylines}
 enter center lat, long of desired tile (q to quit)
43.5 -101.5
ilat,ilon,crustal type:   47  79
 topography:   0.829999983    
  layers: vp,vs,rho,bottom
   1.50   0.00   1.02   0.83
   3.81   1.94   0.92   0.83
   2.50   1.07   2.11   0.33
   4.60   2.59   2.46  -0.77
   0.00   0.00   0.00  -0.77
   6.20   3.60   2.76 -12.43
   6.40   3.70   2.81 -24.08
   6.80   3.90   2.91 -36.08
 pn,sn,rho-mantle:    8.14   4.52   3.35
\end{obeylines}
For this latitude and longitude the distance from sea level was $-$17,618 meters and the density
was 2.81 gm$/$cm$^3$

The Shen, Ritzwoller model\cite{Colorado} is a new (2016) density map only of the United states in $1/4\times 1/4$ degree bins of latitude and longitude.  The density map is divided into many more
layers, than the Crustal map. There are more than 50 layers.  

There is also an older map, PEMC\cite{pemc} included for historical reasons.  

\section{Results}

The result tables with the local radius approximation are given below:
\begin{obeylines}
\begin{obeyspaces}
  number  latitude  longitude  distance seaheight    depth
   \  1        41.833    268.272      0.000    228.444     -2.244
   \  2        41.938    268.918     54.379  -5048.751   5310.851
   \  3        42.043    269.563    108.714  -9852.368  10129.269
   \  4        42.148    270.209    163.003 -14184.244  14364.145
   \  5        42.253    270.854    217.240 -18046.264  18360.764
   \  6        42.359    271.500    271.421 -21440.344  21756.344
   \  7        42.464    272.145    325.542 -24368.449  24652.648
   \  8        42.569    272.791    379.599 -26832.572  27128.373
   \  9        42.674    273.436    433.588 -28834.752  29206.652
   10        42.779    274.082    487.504 -30377.055  30720.654
   11        42.884    274.727    541.344 -31461.594  31838.994
   12        42.989    275.373    595.102 -32090.506  32519.906
   13        43.094    276.019    648.776 -32265.973  32706.572
   14        43.200    276.664    702.362 -31990.203  32440.703
   15        43.305    277.310    755.855 -31265.445  31693.746
   16        43.410    277.955    809.251 -30093.979  30513.578
   17        43.515    278.601    862.547 -28478.111  28977.512
   18        43.620    279.246    915.739 -26420.191  26946.592
   19        43.725    279.892    968.823 -23922.588  24466.488
   20        43.830    280.537   1021.795 -20987.715  21628.814
   21        43.936    281.183   1074.652 -17618.004  18252.004
   22        44.041    281.828   1127.390 -13815.924  14566.324
   23        44.146    282.474   1180.005  -9583.969  10398.169
   24        44.251    283.119   1232.494  -4924.664   5860.664
   25        44.356    283.765   1284.852    159.438   1468.962
----------------------------------------------------------------------------------------------
\end{obeyspaces}
\end{obeylines}

\begin{obeylines}
\begin{obeyspaces}
  number depth   seaheight CRUSTALdens. COLdens. PEMCdens.
   \ 1      -2.244     228.444         2.110       2.280       2.720
   \ 2    5310.851   -5048.751       2.740       2.717       2.720
   \ 3   10129.269   -9852.368       2.740       2.761       2.720
   \ 4   14364.145  -14184.244       2.830       2.788       2.720
   \ 5   18360.764  -18046.264       2.830       2.818       2.720
   \ 6   21756.344  -21440.344       2.830       2.840       2.920
   \ 7   24652.648  -24368.449       2.830       2.873       2.920
   \ 8   27128.373  -26832.572       2.830       2.892       2.920
   \ 9   29206.652  -28834.752       2.830       2.912       2.920
  10   30720.654  -30377.055       2.910       2.930       2.920
  11   31838.994  -31461.594       2.920       2.962       2.920
  12   32519.906  -32090.506       2.920       2.961       2.920
  13   32706.572  -32265.973       2.920       2.935       2.920
  14   32440.703  -31990.203       2.920       2.939       2.920
  15   31693.746  -31265.445       2.830       2.920       2.920
  16   30513.578  -30093.979       2.830       2.911       2.920
  17   28977.512  -28478.111       2.830       2.897       2.920
  18   26946.592  -26420.191       2.830       2.881       2.920
  19   24466.488  -23922.588       2.830       2.861       2.920
  20   21628.814  -20987.715       2.830       2.845       2.920
  21   18252.004  -17618.004       2.810       2.831       2.720
  22   14566.324  -13815.924       2.810       2.811       2.720
  23   10398.169   -9583.969        2.760       2.797       2.720
  24    5860.664   -4924.664        2.760       2.777       2.720
  25    1468.962     159.438         2.760       2.721       2.720
\end{obeyspaces}
\end{obeylines}
%
%
\begin{figure}
\includegraphics[height=21.5cm]{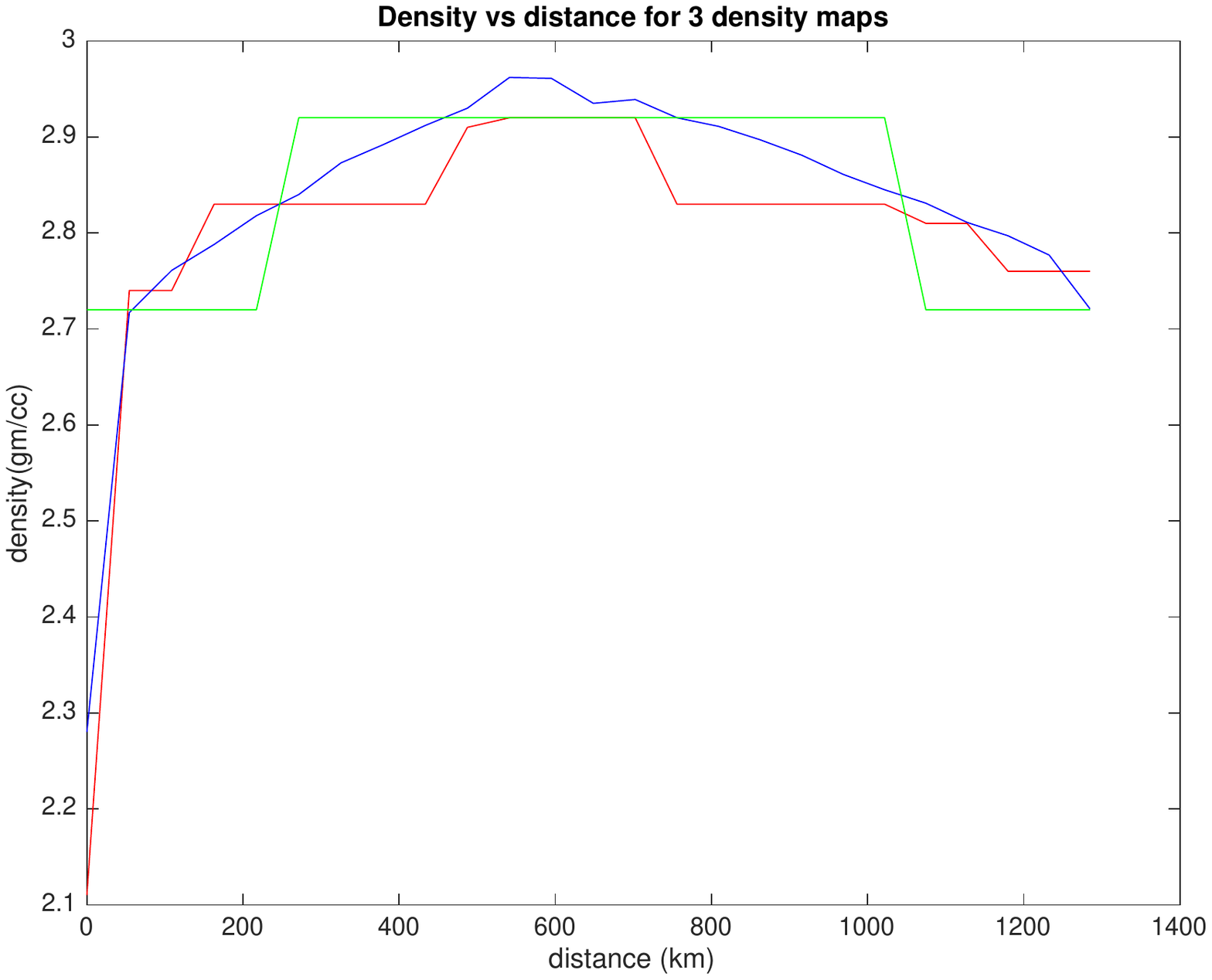}
\caption{ Densities vs distance.  Red is the CRUSTAL map, blue is the Shen-Ritzwoller map,
and green is the old PEM-C map.
}
\label{Figure 4}
\end{figure}
\clearpage
\section{Comments on uncertainties}
Although the actual situation is more complicated, we will look at uncertainties in the total
amount of matter passed through to get an indication of uncertainties.  There are two kinds
of uncertainties to be considered, statistical and systematic.  Statistical uncertainties are due
to random differences.
Sometimes the depths are near a boundary between two densities.
The boundaries are probably not completely flat and there is
some transition region.
Some uncertainties for the crustal map are given below.
\begin{itemize}
\item{} Point 4 has depth $-$14.4 km and crustal has limits for 2.74 at $-$13.10 km
and 2.83 below that.
\item{} Point 8 has depth $-$27.1 km and crustal has limits 
2.83 to $-$27.17 and 2.92 beyond that.
\item{} Point 10 has depth $-$30.7 km and crustal has limits for 2.77 at $-$30.48 km
and 2.92 beyond that.
\item{} Point 11 has depth $-$31.8 km and crustal has limits for of 2.83 down
to $-$30.48 km and 2.92 beyond that.
\item{} Point 13 has depth $-$32.7km and crustal has 2.83 down to $-$31.17 and 
2.92 beyond that.
\item{} Point 14 has depth $-$32.4 km and crustal has 2.83 down to $-$31.17
and 2.92 below that.
\end{itemize}
We have six out of twenty-five path segments with approximately 4\% uncertainties.  If
we view this as a random walk then the standard deviation in the total mount of matter passed through
is 0.43\%.  Even if all twenty-five path segments had a 4\% uncertainty, the 
standard deviation in the total amount of matter passed through would be 0.8\%.  The statistical
uncertainties are quite small.

There are many more layers given for the  Shen,Ritzwoller map
and the differences from layer to layer are of the order of 
1\% (except for the last point, which has 15\% differences).  The statistical uncertainties are
again small.

The systematic uncertainties are those due to a systematic error in the density of the layers.
One approach is to compare the mean density for the three maps:\\
PEMC    2.845 gm/cm$^3$ ;     Crustal    2.817 gm/cm$^3$;     Shen-Ritzwoller;  2.848 gm/cm$^3$. \\
The PEMC map and the Shen-Ritzwoller map have essentially identical means while
the Crustal mean is approximately 1\% lower.  

For the Shen-Ritzwoller map there is another way to estimate errors.  Their density is
calculated from the shear-wave velocity ($vs$) using the empirical formula\cite{Brocher}:
$$\rho = 1.227 +1.53vs-0.837vs^2+0.207vs^3-0.01066vs^4$$
Shen and Ritzwoller are still calculating detailed systematic errors, but they suggest that a reasonable
estimate is to use the standard deviation in $vs$ given in Figure 15 of their publication to
estimate the error in density.  From that figure, 
the standard deviation in the magnitude $vs$ is of the order of 0.03 to 0.05
km/sec over the region of the DUNE beam.  The fractional errors in density 
obtained are fairly constant over the beam path.  
For 0.03, 0.05, and 0.07 km/sec errors in $vs$, one obtains mean fractional errors 
in density of  0.5\%, 0.8\% and 1.2\%.

\section{Acknowledgements}
I wish to acknowledge the considerable help of Professor Henry Pollack of the Earth and 
Environmental Sciences Department, University of Michigan,
in providing considerable expertise to help me understand at least some elementary basics
of the field.

{}

\end{document}